\documentclass[12pt]{article}
\usepackage{graphicx, caption}
\usepackage[utf8]{inputenc}  
\usepackage[T1]{fontenc}
\usepackage{amsfonts,amssymb,amsthm,amsmath}
\usepackage[french,english]{babel}
\usepackage{graphics,graphicx}
        \usepackage{subfigure}                            
\usepackage{amsmath}

\date{}

\begin{document}

\title{Superposition of states in quantum theory}
\author{J.-M.~Vigoureux \\
Institut UTINAM, UMR CNRS 6213,\\
Université Marie et Louis Pasteur,
25030 Besançon Cedex, France\\
jean-marie.vigoureux@univ-.fcomte.fr}                              
\maketitle 
\fontsize{12}{24}\selectfont

\begin{abstract}
The most general mathematical law for summing bounded quantities is not the arithmetic law $+$, but a composition law of which the ‘summation’ law for velocities in special relativity is only one particular example. We believe that this composition law, which we denote by $\oplus$, should also be used in quantum theory. We present it with a few examples, we discuss its physical meaning and we show how it can be useful in quantum theory. We show in particular that its use may open the way for a new interpretation of the use of probabilities in quantum theory.
\newline


\section{introduction}
The standard form for the superposition of two quantum states (for the sake of clarity,we will only consider now the case of two states, but the general case will be considered in the following) is:
\begin{equation}\label{plus states}
|\psi \rangle =c _{1}|\varphi_1 \rangle + \,c_{2}|\varphi_2 \rangle
\end{equation}
In this paper, we present an alternative formulation which we find is fully consistent with measurements to date, and which includes the standard form as a limit. In particular, we consider that the superposition of states should be written as
\begin{equation}\label{comp states}
|\psi \rangle =c _{1}|\varphi_1 \rangle \oplus \,c_{2}|\varphi_2 \rangle =\frac{c _{1}|\varphi_1\rangle + c_{2}|\varphi_2 \rangle}{1 + c _{1}^* \, c_{2}\langle\varphi_1 | \varphi_2 \rangle}
\end{equation}
Les us note that:

\noindent - this expression (\ref{comp states}) reduces to the usual expression (\ref{plus states}) in the (most frequent) case when quantities are small.

\noindent - the denominator of (\ref{comp states}) being a scalar quantity, this "composition law" of states preserves the linearity of the Schrödinger equation.

\noindent - the presence of the denominator in (\ref{comp states}) shows that, in many cases, probabilities calculated using the sum  $+$ (\ref{plus states}) as usual \textit{are found greater than they actually are}.

To illustrate its physical meaning (which we will return to below), let us say briefly that in the particular case of the spatial double-slit experiment, this expression (\ref{comp states}) allows us to express the fact that the particle is not localized before being detected and must therefore be considered as being able to be  
not only at slit $x_1$ \textit{or} at slit $x_2$ but also, as being at slits $x_1$ \textit{and }$x_2$ (a non local property).

This composition law echoes the remarks of R.P. Feynman and P.A.M. Dirac:

\noindent Feynman \cite{Feynman}: "It is a problem of the future to discover the exact manner of computing amplitudes. Today any general law that has been able to deduce from the principle of superposition of amplitudes [...] seems to work. But the detailled interactions still elude us. This suggests that amplitudes will exist in a future theory but their method of calculation may be strange to us."

\noindent Dirac \cite{Dirac}: "Renormalization is just a stop gap procedure. There must be some fundamental changes in our ideas [...] When you get a number turning out to be infinite which ought to be finite, you should admit that there is something wrong with your equations, and no hope that you can get a good theory just by doctoring up that number."

Before turning to the case of quantum theory, we present some applications of this law to show its generality and to show how it generalizes to a superposition of any number of states:

\noindent Its use in the context of special relativity shows us how it generalizes in the complex plane.

\noindent Its use in classical optics shows that using $\oplus$ is equivalent to considering an infinite sum of all possible paths for the particle to go from its source to the detector. Moreover, it shows that this law $\oplus$ involves only \textit{measurable} quantities and not \textit{virtual} quantities so that  the result of summation over all possible \textit{virtual} quantities is directly obtained by composing (using the $\oplus$ law) \textit{measurable} quantities.

We then consider some of its applications in quantum theory, and we examine its physical meaning. 
We show in particular that its use paves the way for an interpretation of quantum probabilities in which \textit{they are not an expression of chance}.
\newline

\indent To introduce the use of this composition law into the general framework of physics, note that
the only natural law to "sum" bounded quantities is the following law which applies the unit disk $D$ of the complex plane on itself : the only automorphisms on $D$ are (apart from a rotation) applications $f$ 
 $$f_a: z  	\longrightarrow a \oplus z=\frac{a + z}{1 + \bar{a} z}$$
So, the \textit{natural} addition law of bounded quantities is not the addition $+$ of arithmetic.
In fact, the use of the addition $+$ can obviously lead to values which fall outside the unit disk and therefore which are not suitable for bounded quantities as are probability amplitudes.

Of course, this work is incomplete and should be developed further, but we're giving here an initial overview.
\section{The example of Special relativity}
To illustrate the use of the composition law $\oplus$  and because some elements will be useful to us later on, let us first consider the composition of velocities in special relativity (note that throughout this section we will take $c =1$).
\subsection{the case of parallel velocities}  
 Consider two reference frames $S$ and $ S'$ moving in the same direction, the velocity of $S'$ with respect to $S$ being denoted $v_1$.  Let us also consider a particle moving in $S'$ in the same direction with the velocity $v_2$. The velocity $V$ of that particle with respect to $S$ is 
 $$V= v_1 \oplus v_2 = \frac{v_1 + v_2}{1 + v_1  v_2}$$
This law is commonly referred to as the relativistic velocity addition formula. Using it:

\noindent - the resultant speed of two velocities with magnitude less than $1$ is always a velocity with magnitude less than $1$.

\noindent - if $v_1$ and $v_2$ are small compared to the speed of light $c=1$,   
then the classical addition law of arithmetic $V = v_1 + v_2$ which is valid in galilean relativity is recovered.

\noindent - this composition law of velocities is the only natural group law to sum real bounded quantities $(v_i<1)$.
 
\indent In the case of a larger number of velocities, the usual calculation may be very long but
the composition law allows us to simply write the result:
 $$V_n = v_1 \oplus v_2 \oplus v_3 \oplus  \text{...} \oplus v_{n-1}\oplus v_n$$ 
whose explicit expression can be written using the elementary symmetrical functions of the polynomials \cite{JMV1991}\cite{JMV1992}:

\indent $\sigma^n_0 = 1$

\indent $\sigma^n_1 = x_1 + x_2 + x_3 + \text{...} x_n $

\indent $ \sigma^n_2 = x_1 x_2 + x_1 x_3 + \text{...}+ x_2 x_3 + \text{...} + x_{n-1} x_n =\sum \limits_{1\leqslant i<j \leqslant n}x_i \, x_j $

\indent ...

\indent $\sigma^n_p = \sum \limits_{1\leqslant i_1 <i_2 <\text{...} <i_p \leqslant n}x_{i_1} \text{...} x_{i_p} $

\indent $\sigma^n_n = x_1\, x_2 \text{...} x_n $

\noindent The result of the relativistic addition of $n$ velocities can thus be written:
$$ V_n=  \frac{\sum \limits_{m \geqslant 0}{} \sigma^n_{2m+1}}{\sum \limits_{m \geqslant 0}{} \sigma^n_{2m}}$$

\noindent For example in the case of three velocities: 
$$ V_3 =  \frac{\sum \limits_{m \geqslant 0}{} \sigma^3_{2m+1}}{\sum \limits_{m \geqslant 0}{} \sigma^3_{2m}} = \frac{v_1 + v_2 + v_3 + v_1 \, v_2 \, v_3}{1+ v_1\,  v_2 + v_2 \, v_3 + v_1\,  v_3}$$
a result which \textit{can directly be obtained} by composing successively $v_1$, $ v_2$ and $v_3$ as: 
$$v_1 \oplus v_2 \oplus v_3 = v_1 \oplus (v_2 \oplus v_3) = v_1 \oplus \dfrac{v_2+v_3}{1+v_2\,v_3}= \text{...}$$
\subsection{the case of non parallel velocities}
In the case of two non-parallel velocities, the usual calculation is rather tedious. 
However, we can easily find the result by generalizing the above composition law in the unit disk of the complex plane \cite{JMV2013}:
let us denote each velocity as $V = v \,e^{ j \theta } $ where $v=||v||$  is the velocity norm and $\theta$ its direction relative to an arbitrary reference axis. 
With this notation, the result of composing two non-colinear velocities can be simply written:
$$V_1 \oplus V_2 = \frac{V_1 + V_2}{1 + \overline{V_1} \, V_2}$$
where $\overline{V}$ denotes the conjugate of $V$ ($\overline{V} = v \,e^{-j \theta} $). 
As above, note that:

- the resultant of two velocities with magnitude less than $1$ is a velocity with magnitude less than $1$.

- if $||v_1||$ and $||v_2||$ are small with respect to the speed of light, then the usual addition $+$ is recovered.

- note also that this law is only \textit{weakly commutative} and \textit{weakly associative} (because of the “bar” in the denominators, $ V_1 \oplus V_2 \ne V_2 \oplus V_1$ although $||V_1 \oplus V_2|| = ||V_2 \oplus V_1||$ so that the above composition law may be said commutative just one phase away).

\noindent This remark is an opportunity to show the interest and the simplicity of using this law $\oplus$: its weakly commutative character in fact directly gives the value of Thomas rotation (which comes from the fact that the composition of two non-collinear boosts results in a boost and a rotation). The weekly commutative character of $\oplus$ in fact gives:
$$ V_1 \oplus V_2 = V_2 \oplus V_1 \, e^{j\, \Theta}$$
where $\Theta$ is the angle of the Thomas rotation \cite{JMV2001,JMV1993a}.

We can simply generalize these results to the case of N coplanar velocities (pay attention to the order of the brackets due to the weak associativity): 
$$ V = v_1 \oplus (v_2 \oplus (v_3 \oplus (\text{...} \oplus (v_{n-1} \oplus v_n))))$$
In the case of \textit{non-coplanar} velocities these results can be generalized using Pauli matrices.
\section{The example of plane parallel plates in Optics}
The probability that a photon will be reflected on a plane parallel plate can be obtained by taking into account all the ways in which it can be reflected by each interface that it encounters in going from the source to the detector that is by summing all the multiple reflections and transmissions at the plate surfaces. Let $r_1 \equiv r_{1,2}$ and $r_2 \equiv r_{2,3}$ be the Fresnel reflection coefficients at the first and at the second interface respectively and let $t_1=t_{1,2}$ and $t'_1=t_{2,1}$ be the transmission coefficient from medium $1$ to medium $2$ and from medium $2$ to medium $1$ respectively.

Taking the incident amplitude equal to unity, and noting that
$r_{2,1}=-r_{1,2}=-r_1$ on can calcultate the overall reflection coefficient $R$ of light on the system by adding all the complex amplitudes of all the possible paths of light inside the plate \cite{BornWolf}:
\begin{equation}\label{paths}
 R = R_1 + t_1\, t'_1\, R_2+ t_1 \,t'_1 \, \overline{R_1}R_2^2 + (-1)^2 t_1 \,t'_1\, \overline{R_1}^2\, R_2^3 + ... + (-1)^n \,t_1\, t'_1 \, \overline{R_1}^n \, R_2^{n+1} +... 
 \end{equation}
 ($R_i = r_i \,e^{j \theta_i}$ with $\theta$ quantifying the phase shift of light when propagating from one interface to the other : $\theta_1$ being the phase when light goes from the light source to the glass and  $\theta$ being that of light between the two surfaces).
 
As in Special relativity, this result can be written simply by using the law $\oplus$ \cite{JMV1993b}:   
$$R = R_1 \oplus R_2 = \frac{R_1 + R_2}{1 + \overline{R_1}\,R_2}$$ 
where $\overline{R_i} = r_i \,e^{-j \theta_i}$. The squared modulus of $R$ represents the probability that a photon will be reflected by the overall system.

\indent This example shows 
that using the composition law $\oplus$ is equivalent to making a sum over the \textit{infinite} number of ways the light could propagate inside the plate (\ref{paths}) (as we do using Feynman path integrals).

In the case of only two interfaces, the result can be obtained simply by noting that the infinite sum over all possible paths of light through the plate is a geometric sequence. This is no longer the case if the number of interfaces is greater than 2. In such cases, even listing all possible paths of light is impossible.
However, as in Special relativity, the result can be easily found by using the $\oplus$ law. Thus, in the case of $n$ interfaces \cite{JMV1993b}: 
$$ R = R_1 \oplus (R_2 \oplus (R_3 \oplus (\text{...} \oplus (R_{n-1} \oplus R_n))))$$
Let's summarize these results : 

- the overall probability amplitude of a photon being reflected by a multilayer is directly given in the most general case by using the composition law $\oplus$.

- using the composition law $\oplus$ is equivalent to doing the \textit{infinite sum on all possible paths of the photon inside the multilayer} exactly as in Feynman path integrals one adds up, or integrates, the amplitudes over the space of all possible paths of the particle between the initial and the final states.

- using the composition law $\oplus$ makes it possible to write the result directly in the case of $n$ interfaces where even the question of simply listing these paths becomes impossible.

- one can add that using the $\oplus$ law 	accelerates the convergence of numerical calculations through a kind of progressive “renormalization”: since the arithmetical addition $R_{n-1}+ R_n$ can lead to a result that is too large, this law immediately divides it by $(1 + \overline{R}_{n-1} R_{n})$ before combining $R_{n-3}$ with the result obtained, and so on. The denominator involved at each step readjusts the result, which would quickly become too large. 

\indent It should be emphasized (this remark is important for the use of $\oplus$ in quantum theory) that this law $\oplus$ involves only \textit{measurable} quantities (the reflection coefficients and phases) and not \textit{virtual} quantities (the infinity of possible virtual paths of light): 
the composition $\oplus$ of the \textit{measurable} quantities is thus \textit{totally equivalent} to the summation over all possible \textit{virtual} optical paths of light in the multilayer. In other words,
the result of summation over all possible \textit{virtual} quantities is directly obtained by using the $\oplus$ law with \textit{measurable} quantities. As a general rule, it is far preferable to use only measurable quantities.
\section{Quantum theory}
The composition law $\oplus$ being the natural law for summing bounded quantities, we may think that it should apply to probability amplitudes. So instead of writing a superposition of states in the form: 
$ |\psi \rangle = c_1 |\varphi_{1}\rangle + c_2 |\varphi_{2}\rangle$  
(with its usual generalization in the case of more than two states)
we should write: 
\begin{equation}\label{TQ}
 |\psi \rangle =c _{1}|\varphi_1 \rangle \oplus \,c_{2}|\varphi_2 \rangle =\frac{c _{1}|\varphi_1\rangle + c_{2}|\varphi_2 \rangle}{1 + c _{1}^* \, c_{2}\langle\varphi_1 | \varphi_2 \rangle}
 \end{equation}
an expression that reduces to the usual one $c_1 |\varphi_1 \rangle +c_2 |\varphi_2\rangle$ in the (\textit{most frequent}) case where $c_{i}|\varphi_i>$ is small.

- Since the denominator of the above law is a scalar quantity, this composition law $\oplus$ also is a solution of the Schrödinger equation.

- The presence of the denominator shows that in many cases probabilities calculated using the sum  $+$ of amplitudes \textit{are greater than they actually are}.

- as in the general case of Special relativity, or in that of wave propagation in a multilayer, the term $c_{1}^*\,c_{2}\,\langle \varphi_1 | \varphi_2 \rangle$ in the denominator reveals that this linear superposition is (in the general case) non commutative ($c _{1}^* \,c _{2}\,\langle \varphi_1 | \varphi_2 \rangle \neq c_{2}^* \,c_{1}\,\langle \varphi_2 | \varphi_1 \rangle)$.

\indent In this connection, we take the liberty of quoting again from Feynman's text, already quoted in the introduction: "It is a problem of the future to discover the exact manner of computing amplitudes. Today any general law that has been able to deduce from the principle of superposition of amplitudes [...] seems to work. But the detailled interactions still elude us. This suggests that amplitudes will exist in a future theory but their method of calculation may be strange to us." \cite{Feynman}.

\subsection{examples of applications}
Our aim now is to show a few examples of the application of (\ref{TQ}). We will then return to its physical meaning.
\newline

\textbf{a) wave functions and energy levels in quantum potential of arbitrary shape} 
 
Wave functions and energy levels in quantum potential of arbitrary shape can be obtained by using $\oplus$. Calculations then require the use of numerical means. We won't return to this question which has already been developed in a previous publication \cite{Grossel 97} but let us note that using the $\oplus$ law, computing is more rapid, more stable, with a remarkable absence of noise. They also show an acceleration in the convergence of the calculation (thanks to the progressive "renormalization" caused by the presence of the denominator of $\oplus$).
Results can be compared with the ones obtained with the WKB method which appears as a first approximation of using $\oplus$.
 \newline
 
 \textbf{b) the hydrogen molecule ion $H_2^+$ }
 
Although this  molecular ion is admittedly of little chemical interest, we present it because it appears to be a good example and also because we will need it in the next part. This ion consists of two hydrogen nuclei sharing a single electron. The lowest energy molecular orbital $\psi^+$ of $H_2^+$ can be obtained by using the LCAO method when noting that in the vicinity of proton $A$, $\psi^+$ must resemble $|\varphi_A \rangle$ (the $|1s_A \rangle$ hydrogen atomic orbital on proton $A$) and in the vicinity of proton $B$,  $\psi^+$ must resemble $|\varphi_B \rangle = |1s_B\rangle$.
The wave function is then obtained by writing the linear superposition:
 $$|\psi\rangle = c_A \, |\varphi_A\rangle +c_B \, |\varphi_B\rangle$$
The two coefficients $c_A$ and $c_B$ are then calculated from energy consideration using the secular equation for $H_2^+$.
 After long calculations one obtains the molecular wavefunction for the bonding molecular orbitals:
$$|\psi^+\rangle =\frac{|\varphi_A\rangle + |\varphi_B\rangle}{\sqrt{2(1 + S)}}$$
with a similar result for the antibonding case. $S= \langle \varphi_A |\varphi_B\rangle$ is a quantitative measure of the overlap of two atomic orbitals $|\varphi_A\rangle$ and $|\varphi_B \rangle$ 
on atoms $A$ and $B$ (note that the overlap integral $S$ is a real quantity since the $|1s \rangle$ orbitals of hydrogen are real functions). 

 This result can be \textit{directly} obtained by writing that the wave function is not an  addition of the two elementary wavefunctions, but their composition by $\oplus$:
$$|\psi\rangle = c \,(|\varphi_A \rangle \oplus |\varphi_B \rangle)= c\, \frac{|\varphi_A \rangle + |\varphi_B\rangle}{1 + \langle \varphi_A |\varphi_B\rangle}$$
 where $c$ is determined by normalization and can easily be found to be $c= \sqrt{\frac{1 + S}{2}}$:
 $$ |c|^2 \frac{||\varphi_A \rangle  + |\varphi_B \rangle|^2}{|1 + \langle \varphi_A |\varphi_B \rangle|^2} =  |c|^2 \frac{2}{1 + S} =1$$
so that we get \textit{directly} the usual result 
$$\psi = \sqrt{\frac{1 + S}{2}}\,\, \frac{|\varphi_A \rangle  + |\varphi_B \rangle }{1 + \langle \varphi_A |\varphi_B \rangle}= \frac{|\varphi_ A \rangle+ |\varphi_B\rangle}{\sqrt{2(1 + S)}}$$
\newline

\textbf{c) comparison of the composition law $\oplus$ and the addition law $+$}

We believe that the natural law of superposition of states is the composition law $\oplus$ and not the usual arithmetic law $+$. We may then wonder why the usual use of addition $+$ leads to such good results.

\noindent This invites us to take a closer look at the law $\oplus$: we have said that it reduces to the usual summation when its terms are “small”. We must therefore ask ourselves what meaning to give to this word.
To do this, the simplest way is to consider interference experiments and, more precisely, to compare the results obtained with a Fabry-Perot interferometer (or with plane parallel plates) with those obtained with Young's slits.

 Such experiments are expected to lead to a summation $+$ of amplitudes in the case of small amplitudes and to a composition $\oplus$ of amplitudes in the case of large amplitudes.
 
\noindent - In the case of plane parallel plates, the reflection coefficients on each interfaces can be close to $1$, and we must therefore expect to find an Airy distribution (corresponding to using the $\oplus$ law). This is indeed what happens.

.\noindent - In the case of Young's slits, amplitudes are obviously much smaller since the slits are necessarily very narrow. We therefore expect to use an addition law $+$. It is indeed what we observe. 
 
The point to emphasize is that the difference between these two results is only clearly visible when the amplitudes are large: it is very difficult to see a difference between the intensity obtained in a Young's slits experiment and that obtained using plane parallel plates when amplitudes are small. So we may think that in every cases the law which should be used is the $\oplus$ law.

\indent Many experiments showing quantum interferences conclude as showing a sine function when in fact they may be showing a curve obtained with $\oplus$ superposition, thereby obscuring the fact that the natural law to be used is not the usual addition but the composition law $\oplus$. 

Figure $1b$ shows a sine function (addition $+$ of amplitudes) and figure $1a$ shows the intensity obtained with a plane parallel plate or with a Fabry-Perot interferometer (composition $\oplus$ of amplitudes) in the case where amplitudes $a=0.2$ (the maximum amplitude being taken as equal to 1). It is clear that it is very difficult to distinguish between the two curves without carefully superimposing them (fig. $2a$).  We can therefore reasonably assume that experiments concluding to a sine function (fig. 1b) and so justifying the usual addition of amplitudes, are in fact showing an Airy distribution (fig. 1a) so showing that a superposition of states must be written with $\oplus$.

These considerations suggest the need for a precise analysis of the interference patterns obtained in experiments.
We therefore closely studied various results found in different publications
to see whether the curves observed were closer to an Airy distribution ($\oplus$ law) than to a sine function ($+$ law), i.e., specifically by looking for a slight asymmetry in the intensity curves observed (even a very small asymmetry would be indicative of the $\oplus$ law).
All studies and analyses show a tendency towards a more or less marked Airy distribution, but these results are not sufficient to draw conclusions, as the study was based on printed figures where accidental distortions may have occurred. It would therefore be interesting to carry out very precise experiments, because, as shown in the fig.$1$, the use of the $\oplus$ law leads to results which can be misinterpreted as sine functions. Observations of the corresponding Airy fringes in a two-slit experiment would be something extremely significant.
\begin{figure}[ht]
\centering\mbox{\subfigure[result obtained using the law $\oplus$]
{\includegraphics[scale=0.3]{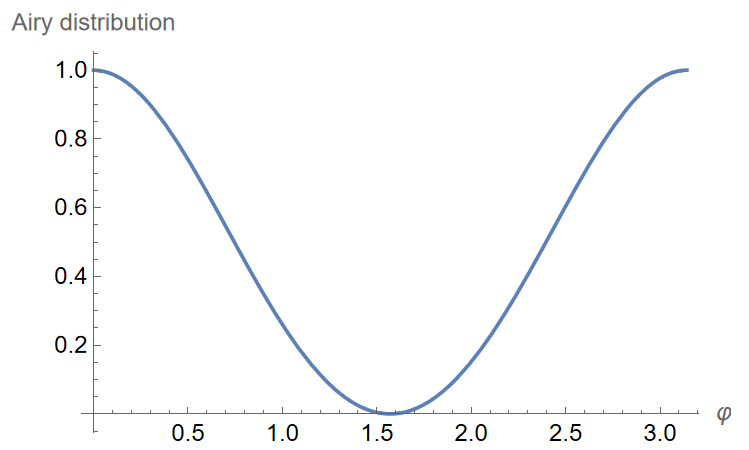}}\quad
\subfigure[result obtained using the law $+$]
{\includegraphics[scale=0.3]{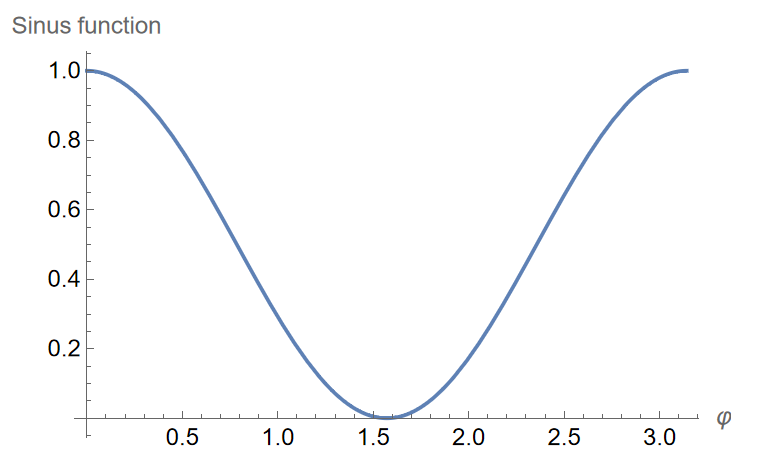}}}
\caption{
Results obtained using amplitude addition $+$ and composition law $\oplus$: the left curve represents the result obtained using amplitude composition law $\oplus$ when $a=0.2$ (the maximum amplitude being taken as $1$). The curve on the right represents the result obtained using the addition law $+$. These results show that, except in the case of large amplitudes, it is very difficult to distinguish between these two laws, the only way being to superimpose them (fig. 2a) or to measure their half-height width very precisely (fig.2a). We can therefore legitimately ask ourselves whether experimental results showing interference are not too quickly presented as showing sinusoids.
 \label{fig1}}
\end{figure}

\begin{figure}[ht]
\centering\mbox{\subfigure[comparison of the results obtained with the addition $+$ of amplitudes and with the composition law $\oplus$ (a\,=\,0.2)] 
{\includegraphics[scale=0.3]{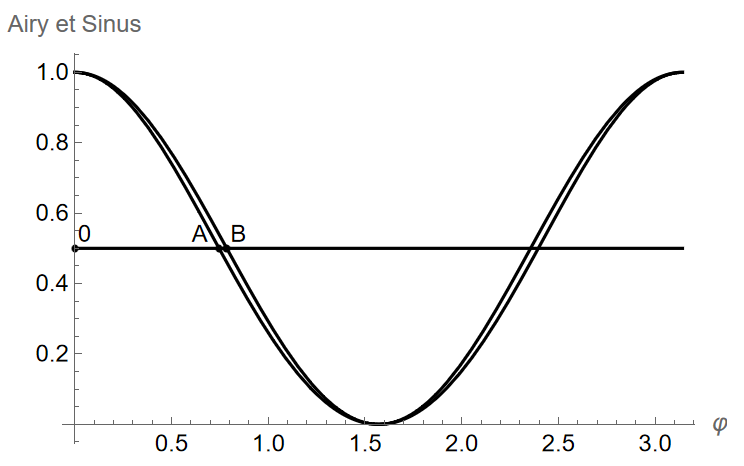}}\quad
\subfigure[difference between the two laws $+$ and $\oplus$ expressed as percentages]
{\includegraphics[scale=0.35]{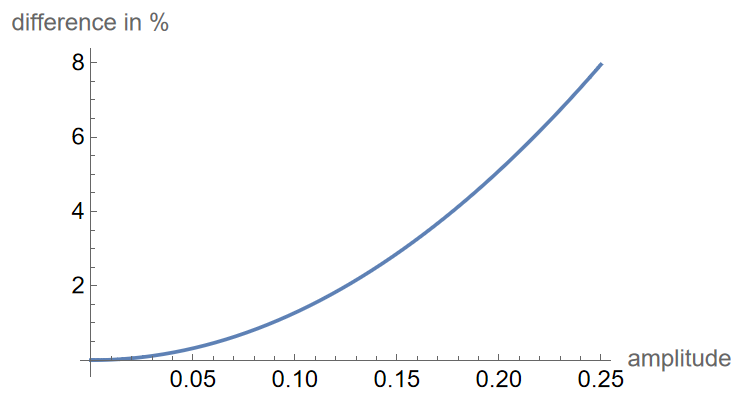}}}
\caption{The comparison between the two laws $+$ and $\oplus$ can be evaluated by comparing the mid-height widths of the corresponding curves, i.e., by comparing $OA$ and $OB$ (fig.$2a$). On the left curves, the amplitudes are equal to $0.2$. The right curve evaluates this difference in percentages: $(OB-OA)/OB$. We can see that up to fairly large amplitude values, the two curves remain difficult to distinguish. So we cannot really conclude that the law to be used in a superposition of states is the $+$ law. \label{fig2}}
\end{figure}
\section{Study of the physical meaning of the denominator of the law $\oplus$}
\indent To discuss the physical meaning of the denominator of the composition law (a question introduced, in a way, when we encountered the overlapping integral), we will use the example of probabilities. Quantum theory uses the laws of classical probabilities, which are simply replaced by similar laws relating to amplitudes (we mean: partial amplitudes rather than probabilities are added, and independant amplitudes rather than probabilities are multiplied). We can therefore rely on classical probabilities to \textit{formally} seek the meaning of the denominator of (\ref{comp states}).

We will show that (\ref{comp states}) expresses a certain ‘fuzziness’ (a ‘non-localisation’ or ‘ubiquity’ in the case of positions) that is overlooked in the calculation of arithmetic sums.

Let us denote $P(A)$ and $P(B)$ the probability of occurences $A$ and $B$ (represented by sets $A$
 and $B$). If $A$ and $B$ are two (eventually) non-mutually exclusive occurences, and if $P_{A\cup B} = 1$ (we will consider the case $P_{A\cup B} \neq 1$ further) 
then
\begin{equation}\label{proba}
P_{A\cup B} = P_A  + P_B  - P_{A\cap B}  = 1  \qquad  \qquad	\Longrightarrow \qquad \qquad  \frac{P_A + P_B}{1+ P_{A\cap B}} =1 
\end{equation}
where we recognize the form of the composition law $\oplus$.

This relationship puts us on the path to interpreting the denominator of the composition law: $P_{A\cap B}$ is the joint probability of A and B. $P_{A\cap B} \neq 0$ shows that $A$ and $B$ can both occur (as the value $<\varphi_1\big| \varphi_2> \neq 0$ in the exchange integral in $H_2^+$ molecular orbitals expresses the fact that the electron can be on the two shells).

We propose to generalize this result by interpreting the denominator of the law $ \oplus$ as expressing the quantum possibility for the particle of being, \textit{for example}, both “here” and “there”
$$P(here \cup there) = P(here) + P(there)  - P(here \cap there) = 1$$
thus leading to a normalization law of the form (\ref{proba}).

So, we must take into account that, \textit{before measurements, the quantities studied may not be mutually exclusive}. Their exclusive nature (for the observer, the particle is never both “here” and “there”) only comes into play at the moment of measurement.

\noindent Two locations \textit{for example} may appear different to us, but there is no evidence that they are different for a quantum particle. As in Feynman's path integrals, the electron is detected at a specific location in the observer's space, but before its detection it can be considered to be in all places and at all times, on all possible paths.

\noindent - The denominator of $\oplus$ thus expresses and reveals a certain ‘fuzziness’ regarding the location in the case of space variables (and more generally regarding the quantity studied).

In the general case where $P_{A \cup B} \neq 0$:
$$P_{A \cup B }+ P_{\subset (A \cup B)} = 1 \qquad  \Rightarrow \qquad P_A  + P_B  - P_{A \cap B}+ P_{\subset (A \cup B) } = 1   $$
$$ \Rightarrow \,\,\,\,  \frac{P_A  + P_B}{1 + P_{A \cap B}} + \frac{P_{ \subset (A\cup B)}}{1 + P_{A \cap B}}=1$$
where we find a relationship similar to the relationship linking the reflection and transmission coefficients $R + T = 1$ in the case of plane parallel plates.

As in the cases considered above (special relativity, optics), these results can be generalized to the case of any number of occurences. In the case of three :
$$ P_{A \cup B \cup C}  = P_A  + P_B + P_C  - P_{A \cap B} - P_{B\cap C} - P_{C\cap A} +   P_{A \cap B \cap C}   $$          
leads to 
$$\frac{P_A + P_B + P_C + P_{A \cap B \cap C } }{1 + P_{A \cap B} + P_{B\cap C} + P_{C\cap A}}=1$$
a result that we can compare to the one we obtain in the case of reflection of light by three interfaces \cite{JMV1992}
$$R=\frac{R_1 + R_2 + R_3 + R_1 \overline{R_2} R_3}{1 + \overline{R_1} R_2 + \overline{R_2} R_3 + \overline{R_1} R3}$$
Just as the term $P_{A \cap B}$ expresses the possibility of seeing the event occur both according to A and according to B, the term $R_i \,R_j$ expresses the possibility of the photon being reflected  \textit{both} by interfaces $i$ and $j$.

This result is important because it shows that, in order to obtain the expected result directly, it is sufficient to take into account the fact that the photon can be reflected  not only by the first \textit{or} second interface, but that it can be also reflected \textit{actually} \textit{by both interfaces}. Thus we can get the correct answer for the probability of reflection of the photon
by taking into account that the photon can get to the detector reflecting off either the front surface or the back surface of the sheet of glass or by the two interfaces at the time.

Thus: the infinite sum over all paths that the photon can take in the plane parallel plate is strictly equivalent to considering that the photon has the possibility of being reflected \textit{actually} by both interfaces.
We see the emergence of a notion of non-localisation (or, when it comes to space variables, a kind of ubiquity of the particle) expressed by the denominator of the composition law $\oplus$.

\section{Conclusion. The meaning of probabilities in quantum theory}
In quantum theory, the superposition of states is expressed using the usual addition law $+$ of arithmetic, but, as special relativity first showed, the only possible law for summing bounded quantities is the composition law $\oplus$.

Numerical comparison of these two laws $+$ and $\oplus$ shows that they are only easy to distinguish in the case of large amplitudes (just as the relativistic law of composition of velocities can only be distinguished from the usual law of addition for large velocities).

In most experiments involving superpositions of states, the amplitudes are small. It is therefore logical that the question of using a more general law has not been raised, but there are many reasons to believe that the law to be used is the $\oplus$ law.       
This composition law has many advantages discussed in this article, but an important consequence of these considerations is that the usual calculations in quantum theory, because they use the summation $+$ of amplitudes instead of the composition $\oplus$ of amplitudes, can lead to theoretical results that are \textit{greater than they actually are}.

An interesting question is how to interpret the physical meaning of the denominator of eq.($\ref{comp states}$). An analysis based on the results of classical probabilities suggests that this denominator reveals the possibility for the particle to be \textit{both} in different places (in the case of space variables) or to have different values of a variable (in the case of other variables).
In the case of the position variable, for example, there is no reason to believe that a particle is somewhere in a specific location before it is detected (nor  even that the concept of space-time is appropriate for it).
\newline

These reflections open the door to another interpretation of probabilities in quantum theory:

In quantum theory, the use of probabilities is interpreted in terms of ‘chance’, leading us to think in terms of ‘indeterminism’, but for physicists, any mathematical theory is merely a tool and therefore does not impose any physical interpretation. To give just one example, it is not because tensors were originally introduced to study elasticity that their use in electromagnetism implies the elasticity of the electromagnetic field. 
Similarly, it is not because probabilities were first developed to study games of chance that their use in quantum theory implies or reveals any action of chance. The above considerations thus suggest that the use of probabilities may express the (total or partial) "irrelevance" (“inadequacy”, "non-pertinence"...) of the use of a quantity.
To say that a particle is equiprobable everywhere would mean that the question of its location is irrelevant (or that the notion of space has no meaning for it).

This interpretation thus follows the historical evolution of the interpretation of Heisenberg's ‘uncertainty relations’, which have become ‘indeterminacy relations’: these relations do not express the fact that we do not know where the particle is, but the fact that, in a certain domain, the very notion of position has no meaning for it.
Thus – and this must be generalized for any variable – it is not because we do not know where the particle is that we use the language of probabilities, but because the particle we are talking about is not “localized”; because the quantity we attribute to it is not “relevant” to describe it (or is only partially relevant).
When the question posed to nature is asked in terms of variables that are not relevant (for example, if we ask ‘where’ a particle that is not localized is located), the answer will be expressed mathematically using probability formalism. Since the variable introduced in the question has no meaning, we are forced to write that all values of this variable are probable. In terms of our macroscopic description of the phenomenon, this can only be done by considering that the particle is ‘here’ or “there” or ‘over there’... provided that we take into account in the calculation that it can also be ‘both here and there,’ ‘here and over there,’ ‘there and over there’... thus introducing the composition law $\oplus$ which takes these latter cases into account in its denominator.

To clarify this interpretation, let us take the example of the polarisation of a photon:

In the orthodox interpretation of quantum theory, a photon emitted by an atom has no polarisation \textit{before} passing through a polariser (we cannot talk about polarisation independently of a measuring device). Under these conditions, how can we calculate the result of its passage through a polarizer? Since it has no defined polarisation (the question of its polarisation before passing through a polariser is irrelevant), we will consider that it has all polarizations and integrate over all its possible orientations (which it does not have) to calculate the result of its passage through the polariser it will encounter.
Knowing that \textit{if} its polarisation direction formed an angle $\theta$ with the polariser axis, its probability of passing through the polariser would be $\cos^2\,\theta$, we write:
\begin{equation}
P = \frac{1}{2 \pi}\ \int_{0}^{2 \pi}\cos^2\,\theta\, d\theta = \frac{1}{2} 
\end{equation}

Thus, since the concept of polarisation is not relevant before the photon passes through a polariser, we are forced to consider that all polarizations are \textit{equally probable} and to sum over all polarizations.

Similarly, the probability amplitude for a particle to go from one space-time coordinate to another, must be calculated by adding up (or integrating) the amplitude over the space of \textit{all possible paths} of the system because the particule is not in space-time and so, the question of its localisation or of its trajectory is not pertinent (the $\oplus$ law takes this fact into account).

 Returning to the example of the position variable, Albert Einstein and Leopold Infeld \cite{Einstein1938} expressed the need to ‘give up describing subatomic particles in space and time’. Similarly, Bohr wrote: ‘the desire for an intuitive representation consistent with images in space and time is not justified’ or ‘no classical idea (wave or particle, causality or space-time) has any validity in atomic physics’.
 
Thus, the use of probability in quantum theory would not express indeterminism (or chance) but rather the \textit{partial or total} irrelevance of the variable chosen to describe the system. This \textit{partial or total inadequacy} is expressed by the law $\oplus$.

It is not because we do not know where the particle is located (in the case of a spatial variable) that we use the language of probability, but because the question of its location is not (or is only partially) relevant.

\end{abstract}

\begin{thebibliography}{99}

\bibitem{Feynman} R. P. Feynman and A. R. Hibbs, \textit{Quantum Mechanics and Path integrals} (McGraw-Hill, New York, 2005), p. 24.

\bibitem{Dirac} P. A. M. Dirac, in P. Buckley and F. D. Peat, \textit{A Question of Physics} (University of Toronto Press, 1979), p. 39.

\bibitem{JMV1991} J. M. Vigoureux, ``Polynomial formulation of reflection and transmission by stratified planar surfaces,'' \textit{J. Opt. Soc. Am.}, 1697–1701 (1991).

\bibitem{JMV1992} J. M. Vigoureux, ``Use of Einstein's addition law in studies of reflection by stratified planar structures,'' \textit{J. Opt. Soc. Am.}, 1313–1319 (1992).

\bibitem{JMV2013} J. M. Vigoureux, ``A generalization in the complex plane of the composition law of non-parallel velocities in Special relativity,'' \textit{Eur. J. Phys.}, \textbf{34}, 795–803 (2013).

\bibitem{JMV2001} J. M. Vigoureux, ``Calculations of the Wigner angle,'' \textit{Eur. J. Phys.}, \textbf{22}, 149–155 (2001).

\bibitem{JMV1993a} J. M. Vigoureux, ``The reflection of light by planar stratified media: the groupoïd of amplitudes and a phase Thomas precession,'' \textit{J. Phys. A: Math. Gen.}, \textbf{26}, 385–393 (1993).

\bibitem{BornWolf} M. Born and E. Wolf, \textit{Principles of optics: electromagnetic theory of propagation, interference and diffraction of light} (Pergamon Press, London; New York; Paris, 1959).

\bibitem{JMV1993b} J. M. Vigoureux and Ph. Grossel, ``A relativistic-like presentation of optics in stratified planar media,'' \textit{Am. J. Phys.}, \textbf{61}(8), 707–712 (1993).

\bibitem{Grossel 97} Ph. Grossel, J.M. Vigoureux, Calculation of wave functions and energy levels: application to multiple quantum wells and continuous potentials, Phy. Rev A, \textbf{55}(1), 796-799 (1997).

\bibitem{Einstein1938} A. Einstein and L. Infeld, \textit{Evolution of Physics} (Cambridge University Press, 1938).

\end{thebibliography}
\end{document}